\documentclass[12pt]{amsart}
\usepackage{amsaddr}
\usepackage{amsmath,amsthm,amsfonts}
\usepackage{tensor}
\usepackage{color}
\usepackage{xcolor}
\usepackage{float}
\usepackage[margin=2cm]{geometry}
\usepackage{bbm}

\usepackage{graphicx}        

\usepackage[bottom]{footmisc}

\usepackage{newtxtext}      %

\newtheorem{prop}{Proposition}[section]

\newcommand{\beprop}{\begin{prop}}
\newcommand{\enprop}{\end{prop}}
\newcommand{\bprf}{\begin{proof}}
\newcommand{\eprf}{\end{proof}}

\usepackage[default]{frcursive}

\usepackage[breaklinks,colorlinks,backref]{hyperref}
\hypersetup{
    colorlinks, 
    linktoc=all, 
    linkcolor=red,  
  citecolor=hyptxt,
  urlcolor=blue}
  \hypersetup{
  citebordercolor=,
  filebordercolor=green,
  linkbordercolor=blue
}
\definecolor{hyptxt}{rgb}{0.7, 0.4, 0.9}
\usepackage{bibtopic}

\usepackage[full]{textcomp} 
     \usepackage{fbb} 
     \usepackage[scaled=.95]{cabin}
     \usepackage[varqu,varl]{inconsolata}
     \usepackage[libertine,bigdelims,vvarbb]{newtxmath}
     \usepackage[cal=boondoxo]{mathalfa}
     \usepackage[T1]{fontenc}
\useosf

\begin{document}

\title[Classical or quantum ?]{\small Signal analysis and quantum formalism: \\ Quantizations with no Planck constant}
\author{\small Jean Pierre Gazeau${}^{a}$ and C\'elestin Habonimana${}^b$}
\address{${}^a$APC, UMR 7164, Universit\'e de Paris, 75205 Paris, France}
\address{${}^b$ Universit\'e du Burundi and Ecole Normale Sup\'erieure,  Bujumbura,  Burundi,} 
\email{gazeau@apc.in2p3.fr, habcelestin@yahoo.fr}
\date{\today}

\begin{abstract}
Signal analysis is built upon various resolutions of the identity in signal vector spaces, e.g. Fourier, Gabor, wavelets, etc. Similar resolutions are used as quantizers of functions or distributions, paving the way to a time-frequency or time-scale quantum formalism and revealing interesting or unexpected features. Extensions to classical electromagnetism  viewed as a quantum theory for waves and not for photons are mentioned.
\end{abstract}


\maketitle
\tableofcontents

\section{Introduction:  the Planck constant $\hbar$ is for what?}
\label{JPGsec:introduction}

Around 1861  Maxwell derived theoretically    the equations that bear his name by using a molecular vortex model of Michael Faraday.  He was certainly not  aware that their content combines  two uppermost physical theories of the next century, namely special relativity and quantum mechanics, even though he was inspired by the relationship  among electricity, magnetism, and the speed of light  (Weber and Kohlrausch, 1855),  $c=1/\sqrt{\epsilon_0\mu_0}$. 
To see that, let us Fourier transform  the second-order Maxwell equations in the vacuum
\begin{equation}
\label{JPGMEST}
\left(\frac{\partial^2}{c^2\partial t^2}- \frac{\partial^2}{\partial x^2}-\frac{\partial^2}{\partial y^2}-\frac{\partial^2}{\partial z^2}\right)\psi(t,\mathbf{ r})= 0\,,\quad \mathbf{r}= (x,y,z)\,,
\end{equation}
where $\psi$ stands generically for components  of magnetic or electric fields or potentials.
We obtain the following equation understood in the sense of distributions,
\begin{equation}
\label{JPGMEFV}
\left(\omega^2 - c^2\,\mathbf{k}^2\right)\, \widehat{\psi} (\omega, \mathbf{ k})=0\,, \quad \widehat{\psi} (\omega, \mathbf{ k})=\frac{1}{4\pi^2}\int_{\mathbb{R}^4}e^{-\mathrm{i}(\omega t-\mathbf{ k}\cdot \mathbf{ x})}\,\psi(t,\mathbf{r})\,\mathrm{d}t\,\mathrm{d}^3\mathbf{ x}\,,
\end{equation}
where $\omega$ is the angular frequency (in radians per second), and $\mathbf{k} = (k_x, k_y, k_z)$, $\mathbf{k}^2:=\mathbf{k} \cdot\mathbf{k}$, is the wave vector (in radians per meter).
One can read \eqref{JPGMEFV} as the quadratic relation  between multiplication frequency and wave vector operators acting in Fourier representation, 
\begin{equation}
\label{JPGKG1}
\left(\Omega^2 -c^2\, \mathbf{K}^2\right)\widehat{\psi} (\omega, \mathbf{k})=0\, , \quad \left(\Omega\,,\mathbf{ K}\right)\,\widehat{\psi} (\omega, \mathbf{ k})= (\omega, \mathbf{k})\,\widehat{\psi} (\omega, \mathbf{ k})\,,
\end{equation}
and, equivalently, in space-time coordinates,
\begin{equation}
\label{JPGKG2}
\Omega\,\psi(t,\mathbf{r}) = -\mathrm{i}\frac{\partial}{\partial t}\,\psi(t,\mathbf{r})\, , \quad \mathbf{K}\,\psi(t,\mathbf{r}) = -\mathrm{i}\nabla_{\mathbf{ r}}\,\psi(t,\mathbf{r})\,.
\end{equation}
Hence, we can see in these apparently trivial manipulations  two ``modern'' aspects of the so-called classical electromagnetism formulated by Maxwell:
\begin{enumerate}
  \item[(i)] the underlying quantisation of frequency and wave vector, $\omega \mapsto \Omega$, $\mathbf{ k}\mapsto \mathbf{ K}$, 
  \item[(ii)] the underlying invariant equation of relativistic dynamics,  $\omega^2 - c^2\mathbf{k}^2=0$.
  \end{enumerate}
The second one can be viewed as a particular case of  the Einstein energy-momentum  equation $E^2-c^2\mathbf{p}^2= m^2c^4$ with $m=0$ (massless particle) after insertion of one proportionality constant allowing to write $E\propto \omega$,  $\mathbf{p}\propto \mathbf{k}$. This constant should have the dimension $[\mathrm{M}\mathrm{L}^2\mathrm{\mathrm{T}^{-1}}]$ of an action. It is precisely the  constant $\hbar$ introduced in 1900 by Planck to explain properly  the electromagnetic radiation emitted by a black body. The 
relations $E=\hbar \omega$,  $\mathbf{p}=\hbar \mathbf{k}$, were subsequently proposed by de Broglie in his 1924 thesis to account for wave-particle duality.  

Hence, we can discern the role of the Planck constant in bridging two worlds, namely classical electromagnetism with phase space $$\{(\mbox{time-space}, \mbox{frequency-wave vector}) \in \mathbb{R}^8\}$$ (no rest mass here) and quantum mechanics built from the classical phase space $$\{(\mbox{space}, \mbox{momentum}) \in \mathbb{R}^6\}\,,$$ and the heuristic quantization rules $\mathbf{r} \mapsto \mathbf{R}$ and $\mathbf{p} \mapsto \mathbf{P}$ (respectively multiplication and derivation operators in space representation), and yielding the well-known  wave equations like Schr\"odinger, Klein-Gordon, Dirac... However there is a trouble concerning the pair (time-energy) due to the lack of existence of a consistent time operator  in quantum mechanics (see for instance the review \cite{timeQM}): according to Pauli's argument \cite{pauli58}, there is no self-adjoint time operator canonically conjugating to a Hamiltonian if the Hamiltonian spectrum is bounded from below. However, we have just displayed the quantum nature of Maxwell equations, which look like the Klein-Gordon equation with no $\hbar$. Restoring the latter as a global factor  would appear as   purely artificial since there  is precisely no mass! Note that at the meeting \textit{World Metrology Day} held on 20  May 2019 it was definitely decided to anchor the S.I. standard of mass, the  kilogram, to    the Planck constant whose value is henceforth fixed to $h=2\pi\hbar =  6.62607015\times10^{-34}\,\mathrm{kg}\, \mathrm{m}^2\,\mathrm{s}^{-1}$. 

Elementary solutions of the Maxwell equations are those  Fourier exponentials (basic wave planes) used to implement the Fourier transform in \eqref{JPGMEFV}. We will see that there is no sound argument preventing the existence of a time operator. Thus, what is really the status of frequency, time, wave vector, ..., in electromagnetism? Classical observables? Quantum observables?

\begin{figure}[h]
\begin{center}
	\setlength{\unitlength}{0.1cm} 
\begin{picture}(140,60)
\put(0,30){\vector(1,0){60}} 
\put(65,30){\vector(1,0){50}} 
\put(30,10){\vector(0,1){45}}
\put(90,30){\vector(0,1){25}}
\put(30,30){\vector(1,1){15}} 
\put(90,30){\vector(1,1){15}} 
\put(27, 27){\makebox(0,0){$O$}}
\put(87, 27){\makebox(0,0){$O$}} 
\put(60, 27){\makebox(0,0){$t$}}
\put(115, 27){\makebox(0,0){$t$}}
\put(27, 55){\makebox(0,0){$\omega$}} 
\put(87, 55){\makebox(0,0){$a$}} 
\put(49, 43){\makebox(0,0){$(t,\omega)$}}
\put(109, 43){\makebox(0,0){$(t,a)$}}
 
\put(45, 45){\makebox(0,0){$\bullet$}}
\put(105, 45){\makebox(0,0){$\bullet$}} 
\end{picture}
\caption{Left: Time-frequency plane $\sim$ Weyl-Heisenberg (Classical observable $f(t,\omega)$). Right: Time-scale half-plane $\sim$ Affine group (Classical observable $f(t,a)$).}
	\label{JPGfig:1}       
	\end{center}
\end{figure}
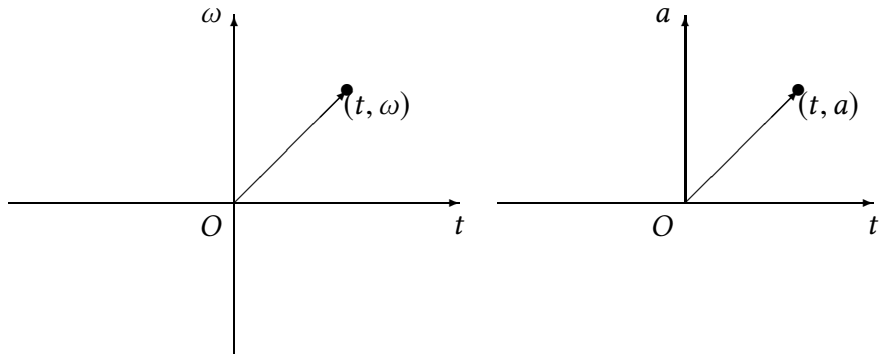

Now, waves in classical electromagnetism are signal, in the true sense of the latter.  This is the leitmotiv of the present contribution, in which we revisit signal analysis where physical quantities are just time and frequency, or time and scale, as they are illustrated by the two phase spaces in Figure \ref{JPGfig:1}.

The organisation of this contribution is as follows. Section \ref{JPGfogawa} is a brief overview of the basic methods in Signal Analysis, namely Fourier, Gabor, and Wavelet.  In Section \ref{JPGsigtoqu} we explain the relationship  between  signal analysis and quantum formalism by pointing out their common Hilbertian framework and the existence in both cases of the essential resolution of the identity. We then define what we mean by quantisation and semi-classical portrait together with their probabilistic content and a possible classical limit, both resulting from a given resolution of the identity. 
In Section \ref{JPGpviq} we implement our approach to quantizations with projector-valued measures provided by Fourier or Dirac bases. They are trivially equivalent to the respective spectral decompositions of the  self-adjoint time and frequency operators, but not allow quantisations of functions of time and frequency.  We enter the heart of our aims and results in Sections \ref{JPGwhiqg} and \ref{JPGaffiq} with resolutions of the identity resulting from the Weyl-Heisenberg and affine groups, and weight functions on the plane or half-plane respectively. The first one stands for the translational symmetry of the time-frequency plane, while the second one stands for the translation-dilation symmetry of the time-scale half-plane (Fig. \ref{JPGfig:1}). Each  one  has a unitary irreducible representation whose square integrability allows to  establish resolution of the identities through Schur's Lemma  and implement the corresponding covariant integral quantisations.  We insist on the fact that whatever the choice of the weight function, time and frequency or scale operators remain essentially the same. In Section \ref{JPGopsa} we illustrate the previous material with examples of operators acting on signals and built from functions $f(b,\omega)$ through Gabor quantization. This gives an idea of the wide range of possibilities in signal analysis offered by our procedure. In Section \ref{JPGdiscussion} we discuss some aspects of our results which we consider as open questions. 

Our contribution may appear as somewhat speculative. We hope that it will open the way to new directions not only in Signal Analysis, but also in Physics through the unveiling of some quantum features of the so-called classical physics.    All the results are given without proof. The latter can be found in previous works \cite{JPGber14,JPGbecuro17,JPGmur16,JPGgaz18,JPGbergaz18,JPGbeczuma18,JPGkono19}.

\section{Fourier, Gabor, and wavelet analysis in a nutshell}
\label{JPGfogawa}
In this section we recall the basics of these three types of signal analysis and fix our Hilbertian notations in terms of Dirac \textit{ket(s)}, $|\cdot\rangle$, as vectors in a Hilbert space and \textit{bra(s)},  $\langle\cdot|$, as elements of its dual. A temporal ``finite-energy''  signal $s(t)$ is viewed as a vector denoted by $|s\rangle$, or by the abusive $|s(t)\rangle$, in the Hilbert space $L^2(\mathbb{R}^2, \mathrm{d}t)$. Its energy is precisely $\Vert s\Vert^2= \langle s|s\rangle$.

\subsection{Fourier analysis, like in Euclidean Geometry ... }

Fourier analysis  rests upon the family of elementary signals   $\frac{1}{\sqrt{2\pi}}\, e^{\mathsf{i}\omega t}$, $\omega \in \mathbb{R}$. 
These non square-integrable functions  can be considered as forming a  ``continuous orthonormal basis'' in the following sense
\begin{gather}
\left\langle \frac{1}{\sqrt{2\pi}}\, e^{\mathsf{i}\omega t} \right |\left. \frac{1}{\sqrt{2\pi}}
\, e^{\mathsf{i}\omega' t} \right\rangle = \frac{1}{2\pi}\int\limits_{-\infty}^{\infty} 
e^{\mathsf{i}(\omega' -
\omega) t}\, \mathrm{d} t
= \delta(\omega' - \omega) \quad  \mbox{``orthonormality''}\,.
\end{gather}
\begin{equation}
\mathbbm{1} = \int\limits_{-\infty}^{\infty} \left| \frac{1}{\sqrt{2\pi}}\, e^{\mathsf{i}\omega t} 
\right\rangle \left\langle \frac{1}{\sqrt{2\pi}}\, e^{\mathsf{i}\omega t}  \right| \, \mathrm{d}\omega\quad
\mbox{``continuous basis solving the identity''}\,.
\end{equation}
Then the
inverse Fourier transform is viewed as the  Hilbertian decomposition  in elementary signals :
\begin{equation}
| s(t)\rangle = \int\limits_{-\infty}^{\infty} 
\underset{\mbox{\it \small Fourier transform $\hat s(\omega)= \frac{1}{\sqrt{2\pi}}\int_{-\infty}^{+\infty}e^{-\mathrm{i}\omega t}s(t)\mathrm{d}t$}}{\underbrace{\left\langle \frac{1}{\sqrt{2\pi}}\, e^{\mathsf{i}\omega t} \right| \left.s(t)
\right\rangle }}\left|
\frac{1}{\sqrt{2\pi}}\, e^{\mathsf{i}\omega t}
\right\rangle
\, \mathrm{d}\omega\,, 
\end{equation}
together with the norm or energy conservation (Plancherel)
\begin{align}
 \Vert s \Vert^2  =\int\limits_{-\infty}^{\infty}  \vert s (t) \vert^2\, \mathrm{d} t 
=  \int\limits_{-\infty}^{\infty}  \vert\hat{s} (\omega)\vert^2 
\, \mathrm{d}\omega = \Vert\hat{s} \Vert^2 
\end{align}

\subsection{Gabor Signal Analysis ($\sim$ time-frequency) }

 The ingredients of the Gabor transform, or time-frequency representation, of a signal are  translation  combined with modulation. 

One chooses a probe, or  \textit{window} or \textit{Gaboret}, $\psi$ which is well localized in time  and frequency at once, and which is normalized, $\Vert \psi \Vert = 1$. This probe is then translated in time and frequency, 
but its size is not modified (in modulus):
\begin{equation}
\label{JPGGabFam}
\psi(t) \rightarrow  \psi_{b, \omega}(t) = 
 e^{\mathsf{i}\omega t}\,\psi (t -b)\,.  
\end{equation}

The time-frequency or Gabor transform is then :
\begin{equation}
s(t) \rightarrow S[s](b,\omega ) \equiv S(b,\omega ) = \langle \psi_{b,\omega} | s \rangle = 
\int_{-\infty}^{+ \infty}
e^{-\mathsf{i}\omega t}\,\overline{\psi (t -b)} \, s(t) \, \mathrm{d} t
\end{equation}

It is easy to prove that there is conservation of the energy :
\begin{equation}
\Vert s \Vert^2 = \int_{-\infty}^{+ \infty} \vert s(t) \vert^2 \, dt = 
\int_{-\infty}^{+ \infty}\int_{-\infty}^{+ \infty} \vert S(b,
\omega) \vert^2 \,\frac{\mathrm{d}b \, \mathrm{d}\omega }{2 \pi}\overset{\mathrm{def}}{=} \Vert S \Vert^2\,,
\end{equation}
  and so the \textit{reciprocity} or \textit{reconstruction} formula holds as:
\begin{equation}
 s(t)= 
\int_{-\infty}^{+ \infty}\int_{-\infty}^{+ \infty}  S(b,\omega)  
e^{i\omega t}\psi (t -b)\, \frac{\mathrm{d}b \, \mathrm{d}\omega }{2 \pi}\,.
\end{equation}
This reconstruction holds in the Hilbertian sense. It results from resolution of the identity provided by the continuous non-orthogonal family $\{\psi_{b,\omega}\,,\, (b,\omega)\in \mathbb{R}^2\}$ (overcompleteness):
\begin{equation}
\label{JPGgabres1}
\mathbbm{1} = \int_{-\infty}^{+ \infty}\int_{-\infty}^{+ \infty}\frac{\mathrm{d}b \, \mathrm{d}\omega }{2 \pi} \left|
\psi_{b,\omega}\right\rangle \left\langle \psi_{b,\omega} \right| \,.
\end{equation}

\subsection{Continuous wavelet transform ($\sim$ time-scale)}
\label{JPGTST}

In this analysis, one also picks  a  well localized ``mother wavelet'' or probe $\psi (t) \in L^2(\mathbb{R}, \mathrm{d}t) $, but we impose more on it: its Fourier transform should be  zero at the origin,  which implies zero average for $\psi$, i.e. $\widehat{\psi}( 0) = \frac{1}{\sqrt{2\pi}}\int\limits_{-\infty}^{\infty} \psi (t) \, \mathrm{d} t = 0$, and with modulus even.  Such conditions are encapsulated in: 
\begin{equation}
0< c_{\psi} :=  2\pi\int\limits_{0}^{\infty} \vert \hat{\psi} (\omega) \vert^2\, \frac{\mathrm{d}\omega}{\omega} 
=  \int\limits_{0}^{\infty} \vert \hat{\psi} (-\omega) \vert^2  \,
\frac{\mathrm{d}\omega}{\omega} < \infty \, . 
\end{equation}
 Then we build the (continuous) family of translated-dilated-contracted versions of $\psi$:
\begin{equation}
\left\{\psi_{b,a}(t)=\frac{1}{\sqrt{a}} \psi\left(\frac{t-b}{a}\right) \right\}_{a \in \mathbb{R}_+^{\star}, b \in \mathbb{R}}\,.
\end{equation}
We obtain   an
overcomplete family in $L^2(\mathbb{R}, \mathrm{d}t)$, which means that any signal  decomposes as
\begin{equation}
\label{JPGCWT1}
| s(t)\rangle = \frac{1}{c_{\psi}} \int\limits_{-\infty}^{\infty}\mathrm{d} b  \int\limits_{0}^{\infty}\frac{\mathrm{d} a}{a^2} S(b,a) \left|\frac{1}{\sqrt{a}} \psi\left(\frac{t-b}{a}\right)\right\rangle \,.
\end{equation}
The coefficient $S(b,a)$, as a function of the two continuous variables  $b$ (time) and $a$ (scale),
is the continuous wavelet transform of the signal: 
\begin{equation}
\label{JPGCWT2}
S(b,a) = \left\langle \frac{1}{\sqrt{a}} \psi\left(\frac{t-b}{a}\right)  \right| \left. s \right\rangle  = \int\limits_{-\infty}^{\infty}
\frac{1}{\sqrt{a}} \overline{\psi}\left(\frac{t-b}{a}\right)  s(t) \, \mathrm{d} t \,. 
\end{equation}
These equations derive from the resolution of the identity  provided by the non orthogonal $|\psi_{ba}\rangle$'s: 
\begin{equation}
\label{JPGCWT3}
\mathbbm{1} =  \frac{1}{c_{\psi}} \int\limits_{-\infty}^{\infty}\mathrm{d} b  \int\limits_{0}^{\infty}\frac{\mathrm{d} a}{a^2}\left | \frac{1}{\sqrt{a}} \psi\left(\frac{t-b}{a}\right)\right\rangle \left\langle
\frac{1}{\sqrt{a}}
\psi\left(\frac{t-b}{a}\right)\right |\,.
\end{equation}
 Energy conservation holds as well, but its repartition in the half-plane stands  with respect to the Lobatchevskian geometry determined by the  measure $\mathrm{d} b\,\mathrm{d} a/a^2$. The latter is left-invariant  under the affine transformations $(b,a) \mapsto (b^{\prime},a^{\prime})(b,a)= (b^{\prime} +a^{\prime}b, a^{\prime}a)$. 
\begin{equation}
\Vert s \Vert^2=  \frac{1}{c_{\psi}} \int\limits_{-\infty}^{\infty}\mathrm{d} b  \int\limits_{0}^{\infty}\frac{\mathrm{d} a}{a^2} \vert S(b,a)\vert^2\, , \quad c_{\psi} \equiv 2\pi\int\limits_{0}^{\infty} \vert \hat{\psi} (\omega) \vert^2\, \frac{\mathrm{d}\omega}{\vert \omega \vert}\,. 
\end{equation}

\section{From Signal Analysis to Quantum Formalism}
\label{JPGsigtoqu}

\subsection{Resolution of the identity as the common guideline}

 Given  a measure space $(X,\mu)$ and a (separable) Hilbert space $\mathcal{H}$,  an operator-valued function
\begin{equation*}
\label{}
X\ni x \mapsto \mathsf{M}(x)\ \mbox{acting in} \ \mathcal{H}\, , 
\end{equation*}
resolves the identity operator $\mathbbm{1}$ in $\mathcal{H}$ with respect the measure $\mu$ if 
\begin{equation}
\label{resUn}
\int_X \mathsf{M}(x)\, \mathrm{d} \mu(x)= \mathbbm{1} 
\end{equation}
holds in a weak sense.

  In Signal Analysis,  \textit{analysis} and \textit{reconstruction} are grounded in  the application of \eqref{resUn} on a signal, i.e., a vector in $\mathcal{H}$
\begin{equation*}
\label{ }
\mathcal{H}\ni | s\rangle \overset{\mbox{{\scriptsize reconstruction}}}{=} \int_X \overset{\mbox{{\scriptsize analysis}}}{\overbrace{\mathsf{M}(x)|s\rangle}}\, \mathrm{d} \mu(x)\,.
\end{equation*}
 In quantum formalism, \textit{integral quantization} is grounded in  the linear map of a function on $X$ to an operator in $\mathcal{H}$
\begin{equation*}
\label{ }
f(x) \mapsto \int_X f(x) \mathsf{M}(x)\, \mathrm{d} \mu(x)= A_f\, , \quad 1 \mapsto \mathbbm{1}\,.
\end{equation*}

\subsection{ Probabilistic content of integral quantization: semi-classical portraits}

If the operators $\mathsf{M}(x)$ in
\begin{equation}
\label{resunH}
\int_X \mathsf{M}(x)\, \mathrm{d} \mu(x)= \mathbbm{1}\, ,  
\end{equation}
are nonnegative, i.e., $\langle \phi | \mathsf{M}(x)|\phi\rangle  \geq 0$ for all $x\in X$, one says that  they form a (normalised) positive operator-valued measure (POVM) on $X$. 

  If they are further
 unit trace-class, i.e. $\mathrm{tr}(\mathsf{M}(x)) = 1$ for all $x\in X$, i.e., if the $\mathsf{M}(x)$'s are density operators, then the map 
\begin{equation}
f(x) \mapsto  \check f(x):= \mathrm{tr}(\mathsf{M}(x)A_f) = \int_X f(x^{\prime})\,\mathrm{tr}(\mathsf{M}(x)\mathsf{M}(x^{\prime}))\, \mathrm{d} \mu(x^{\prime})
\end{equation}
is a local averaging of the original $f(x)$ (which can very singular, like a Dirac !) with respect to the probability distribution on $X$,
\begin{equation}
x^{\prime} \mapsto \mathrm{tr}(\mathsf{M}(x)\mathsf{M}(x^{\prime}))\,. 
\end{equation}
This averaging, or semi-classical portrait of the operator $A_f$,  is in general a regularisation, depending of course on the topological nature of the measure space $(X,\mu)$ and the functional properties of the $\mathsf{M}(x)$'s.

\subsection{Classical limit}

 Consider a set of parameters $\boldsymbol{\kappa}$   and corresponding families of  POVM  $ \mathsf{M}_{\boldsymbol{\kappa}}(x)$ solving the identity
\begin{equation}
\label{resunH}
\int_X \mathsf{M}_{\boldsymbol{\kappa}}(x)\, \mathrm{d} \mu(x)= \mathbbm{1}\, ,  
\end{equation}
 One says that the classical limit $f(x)$ holds at $\boldsymbol{\kappa}_0$ if 
\begin{equation}
\check f_{_{\boldsymbol{\kappa}}}(x):= \int_X f(x^{\prime})\,\mathrm{tr}(\mathsf{M}_{\boldsymbol{\kappa}}(x)\mathsf{M}_{\boldsymbol{\kappa}}(x^{\prime}))\, \mathrm{d} \mu(x^{\prime}) \to f(x) \quad \mbox{as} \quad \boldsymbol{\kappa} \to \boldsymbol{\kappa}_0\, , 
\end{equation}
where the convergence $\check f\to f$ is defined in the sense of a certain topology. 

 Otherwise said, $\mathrm{tr}(\mathsf{M}_{\boldsymbol{\kappa}}(x)\mathsf{M}_{\boldsymbol{\kappa}}(x^{\prime}))$ tends to  
\begin{equation}
\mathrm{tr}(\mathsf{M}_{\boldsymbol{\kappa}}(x)\mathsf{M}_{\boldsymbol{\kappa}}(x^{\prime})) \to \delta_x(x^{\prime})
\end{equation}
where $\delta_x$ is a Dirac measure with respect to $\mu$,
\begin{equation}
\int_X f(x^{\prime}) \, \delta_x(x^{\prime})\, \mathrm{d} \mu(x^{\prime}) = f(x)\,.
\end{equation}
 Of course, these definitions should be given a rigorous mathematical sense, and nothing guarantees the existence of such  a limit.

\section{First examples of operators $M(x)$: projector-valued (PV) measures  for signal analysis}
\label{JPGpviq}
 The measure space is $(\mathbb{R}, \mathrm{d} x)$, and the Hilbert space is $L^2(\mathbb{R},\mathrm{d}x)$. Variable $x$ represents time, $x=t$, or represents frequency, $x=\omega$. The identity is solved by two types of continuous ``orthogonal bases'' 
\begin{enumerate}
  \item Dirac basis $|\delta_t\rangle \equiv |t\rangle$  for time analysis (trivial sampling), based on the well-known $\int_{-\infty}^{+\infty}\delta(t-t^{\prime})\, f(t^{\prime})\, \mathrm{d} t^{\prime} = f(t)$,
  \begin{equation}
\label{ }
\{ \delta_t\, , \, t\in \mathbb{R}\}\,, \quad \langle \delta_t|\delta_{t^{\prime}}\rangle= \delta(t-t^{\prime})\,, \quad  \int_{-\infty}^{+\infty} |\delta_t\rangle\langle \delta_t|\, \mathrm{d} t= \mathbbm{1}\, ,
\end{equation}
with resulting analysis-reconstruction
  \begin{equation}
\label{ }s(t)= \int_{-\infty}^{+\infty}\delta(t-t^{\prime})\, s(t^{\prime})\, \mathrm{d} t^{\prime} \Leftrightarrow |s\rangle = \int_{-\infty}^{+\infty} |\delta_t\rangle\langle \delta_t|s\rangle \, \mathrm{d} t \, ,\quad \langle \delta_t|s\rangle= s(t)
\end{equation}
  \item Fourier basis $\chi_{\omega}(t) = e^{\mathsf{i} \omega t}/\sqrt{2\pi}$ for frequency analysis
   \begin{equation}
\label{ }
\{\chi_{\omega}\, , \, \omega\in \mathbb{R}\}\,, \quad \langle \chi_{\omega}|\chi_{\omega^{\prime}}\rangle= \delta(\omega-\omega^{\prime})\,, \quad  \int_{-\infty}^{+\infty} |\chi_{\omega}\rangle\langle\chi_{\omega}|\, \mathrm{d} \omega= \mathbbm{1}\, ,
\end{equation}
with resulting Fourier analysis-reconstruction
  \begin{equation}
\label{ } s(t)= \int_{-\infty}^{+\infty}\,\chi_{\omega}(t) \hat s(\omega)\, \mathrm{d} \omega \Leftrightarrow |s\rangle = \int_{-\infty}^{+\infty} |\chi_{\omega}\rangle\langle \chi_{\omega}|s\rangle \, \mathrm{d} \omega\, ,\quad \langle \chi_{\omega} |s\rangle= \hat s(\omega)
\end{equation}
\end{enumerate}
In the next we apply the Fourier PV quantization to the elementary  time and frequency variables.

\subsection{PV measures  for quantization: Time operator}

 The time operator  $T \equiv A_t $ is obtained as 
\begin{equation}
\label{specdecT}
t \mapsto A_t  = T = \int_{-\infty}^{+\infty} t |\delta_t\rangle\langle \delta_t|\, \mathrm{d} t\,.
\end{equation}
 From 
\begin{equation*}
|s^{\prime}\rangle:=  T|s\rangle = \int_{-\infty}^{+\infty} t |\delta_t\rangle \underset{s(t)}{\underbrace{\langle\delta_t|s\rangle}}\, \mathrm{d} t\, , \quad \langle\delta_t|s^{\prime}\rangle = t s(t)\,.
\end{equation*}
one sees that $T$ is the multiplication operator $(Ts)(t)= ts(t)$
This operator, with domain the Schwartz space $\mathcal{S}$,   is \textit{essentially self-adjoint}  in $L^2(\mathbb{R},\mathrm{d}x)$ and \eqref{specdecT} is nothing but its spectral decomposition.

\subsection{PV measures  for quantization: Frequency operator}

In turn, the frequency operator $\Omega$ is 
\begin{equation}
\label{specdecOm}
\omega \mapsto A_{\omega} \equiv \Omega = \int_{-\infty}^{+\infty} \omega\,  |\chi_{\omega}\rangle\langle \chi_{\omega}|s\rangle \, \mathrm{d} \omega\,.
\end{equation}
 From
\begin{align}
|s^{\prime}\rangle &:=  \Omega|s\rangle = \int_{-\infty}^{+\infty} \omega |\chi_{\omega}\rangle \underset{\hat s(\omega)}{\underbrace{\langle\chi_{\omega}|s\rangle}}\, \mathrm{d} \omega\, , \\ \langle\delta_t|s^{\prime}\rangle &= \int_{-\infty}^{+\infty} \omega\,\hat s(\omega)\, \frac{e^{\mathsf{i} \omega t}}{\sqrt{2\pi}}\,\mathrm{d} \omega= -\mathsf{i}\, \partial_{t} \int_{-\infty}^{+\infty}\hat s(\omega)\, \frac{e^{\mathsf{i} \omega t}}{\sqrt{2\pi}}\,\mathrm{d} \omega\,.
\end{align}
it is the derivation operator $(\Omega s)(t)= -\mathsf{i} \,\partial_{t}s(t)$. Symmetrically to the previous case, this operator, with domain the Schwartz space $\mathcal{S}$,   is \textit{essentially self-adjoint}  in $L^2(\mathbb{R},\mathrm{d}x)$ and \eqref{specdecT} is nothing but its spectral decomposition.

\subsection{Mutatis Mutandis ... and CCR}

 In a symmetrical way we can write:
\begin{align}
\label{ }
\Omega &=  \int_{-\infty}^{+\infty} \omega |\delta_{\omega}\,\rangle\langle\delta_{\omega}|\, \mathrm{d} \omega\, \quad (\Omega \hat s)(\omega) = \omega \,\hat s(\omega)\,,\\ T&= \int_{-\infty}^{+\infty}t\, |\chi_t\rangle\langle\chi_t|\, \mathrm{d} t\, , \quad (T \hat s)(\omega) = \mathsf{i} \partial_{\omega} \hat s(\omega)\,.
\end{align}
Time and frequency operators obey the ``canonical" commutation rule (a CCR with no $\hbar$ !)
\begin{equation}
T\,\Omega -\Omega\,T \equiv [T,\Omega] = \mathsf{i} \,I\,,
\end{equation}
with its immediate Fourier uncertainty consequence 
\begin{equation}
\label{ }
\Delta_s T\, \Delta_s \Omega\geq \frac{1}{2}\, , \quad \Delta_s A:= \sqrt{\langle s| A^2|s\rangle  - (\langle s|A|s\rangle)^2}
\end{equation}
  Now, one should keep in mind that CCR  $[A,B] = \mathrm{i} \mathbbm{1}$  for a  self-adjoint ($A$, $B$) pair, with common domain, holds true only if both have continuous spectrum $(-\infty,+\infty)$. The expression of the CCR in terms of the respective unitary operators reads as
  \begin{equation}
\label{JPGWeylrel}
 e^{\mathrm{i} \sigma \Omega}\,e^{\mathrm{i} \tau T}= e^{\mathrm{i}\sigma \tau} e^{\mathrm{i}\tau T}\, e^{\mathrm{i} \sigma \Omega}\,,  \quad \mbox{(Weyl relations)}\,.
\end{equation}
 von Neumann proved (1931) \cite{JPGvonneumann31,JPGvonneumann32} that up to multiplicity and unitary equivalence the Weyl relations have only one solution (see \cite{JPGreedsimon75} for the proof).
 
 \subsection{ Limitations of  PV quantization}

The previous quantizations based on spectral PV measures have clearly very limited scopes. As a matter of fact,  they are just equivalent to the spectral decomposition of functions of $t$ or of functions of $\omega$: 
\begin{align*}
\label{ }
f(t) &\mapsto A_f  = \int_{-\infty}^{+\infty} \mathrm{d} t\, f(t)\,\left\lbrace\begin{array}{cc}
  |\delta_t\rangle\langle\delta_t|     &  \mbox{\small multiplication operator on}\ L^2(\mathbb{R},\mathrm{d} t)  \\
    |\chi_t\rangle\langle\chi_t|   &   \mbox{\small (pseudo-)differential operator on} \ L^2(\mathbb{R},\mathrm{d} \omega)
\end{array}  \right.  \\ 
\hat f(\omega)& \mapsto A_f= \int_{-\infty}^{+\infty} \mathrm{d} \omega \, \hat f(\omega) \left\lbrace\begin{array}{cc}
  |\delta_\omega\rangle\langle\delta_\omega |     &  \mbox{\small multiplication operator on}  \ L^2(\mathbb{R},\mathrm{d} \omega)\\
    |\chi_\omega\rangle\langle\chi_\omega|   &   \mbox{\small (pseudo-)differential operator on} \ L^2(\mathbb{R},\mathrm{d} t)
\end{array}  \right.  
\end{align*}
So, the question is how to manage time-frequency functions $f(t,\omega)$?

\section{Weyl-Heisenberg covariant  integral  quantization with Gabor and beyond}
\label{JPGwhiqg}

\subsection{From PV quantization to Gabor POVM quantization}

In order to manage time-frequency functions $f(b,\omega)$, we naturally  think to the resolution of the identity provided by the Gabor POVM introduced in \eqref{JPGgabres1} that we remind below.
\begin{equation*}
\label{ }
\mathbbm{1}= \int_{\mathbb{R}^2}\frac{\mathrm{d} b\,\mathrm{d} \omega}{2\pi} |\psi_{b \omega}\rangle \langle \psi_{b \omega}\rangle| \, ,\quad  \langle \psi_{b \omega}| \psi_{b^{\prime} \omega^{\prime}}\rangle\neq \delta(b-b^{\prime})\, \delta(\omega-\omega^{\prime})
\end{equation*}
 where $\langle\delta_t| \psi_{b \omega}\rangle= e^{\mathsf{i} \omega t} \psi(t-b)$ are the modulated-transported unit-norm probe-vectors in $L^2(\mathbb{R},\mathrm{d} t)$
and where $\mathbb{R} \supset\Delta \mapsto  \int_{\Delta}\frac{\mathrm{d} b\,\mathrm{d} \omega}{2\pi} |\psi_{b \omega}\rangle \langle \psi_{b \omega}\rangle|$ is the corresponding normalised positive operator-valued measure on the plane. Then 
the quantization of $f(b,\omega)$ is given by:
\begin{equation}
\label{JPGgabq1}
f\mapsto A_f = \int_{\mathbb{R}^2}\frac{\mathrm{d} b\,\mathrm{d} \omega}{2\pi}\, f(b,\omega)\,  |\psi_{b \omega}\rangle \langle \psi_{b \omega}\rangle| \, .
\end{equation}
The corresponding semi-classical portrait is given by
\begin{equation}
\label{JPGgabsc1} \check f(b,\omega)=  \int_{\mathbb{R}^2}\frac{\mathrm{d} b^{\prime}\,\mathrm{d} \omega^{\prime}}{2\pi}\, f(b^{\prime},\omega^{\prime})\,  \vert\langle \psi_{b \omega}|\psi_{b^{\prime} \omega^{\prime}}\rangle\vert^2
\end{equation}
Applied to the time and frequency variables, we find that nothing basic is lost with regard to the Fourier PV quantization:
 \begin{align}
\label{JPGgabqt}
    A_b &= T + \mathrm{Cst}_1\mathbbm{1} \,,  \\
 \label{JPGgabqf}     A_\omega &= \Omega + \mathrm{Cst}_2\mathbbm{1} \,.
\end{align}  
where the additive real constants are easily cancelled through an appropriate choice of $\psi$.

\subsection{Beyond Gabor quantization}

Gabor signal analysis and quantization are the simplest ones among a world of possibilities, all of them being based on the unitary dual of the Weyl-Heisenberg group. 
Let us remind the most important  features of this group  that we use in our approach to quantization. More details are given in the pedagogical 
\cite{JPGgaz18}. 

We recognize in the construction of the Gabor  family \eqref{JPGGabFam} the combined actions of the two unitary operators introduced in \eqref{JPGWeylrel}, with respective generators the self-adjoint time and frequency operators
\begin{equation}
\label{JPGGabWH1}
L^2(\mathbb{R}, \mathrm{d}t) \ni \psi(t) \mapsto  \psi_{b,\omega}(t)=  \left(e^{\mathrm{i} \omega T} e^{-\mathrm{i} b\Omega} \psi\right)(t)\, ,
\end{equation}
Two alternative forms of the action \eqref{JPGGabWH1} are provided by the Weyl formulae \eqref{JPGWeylrel} combined with the Baker-Campbell-Hausdorff formula:
\begin{equation}
\label{JPGGabWH2}
\psi_{b,\omega}(t) =e^{\mathrm{i} b\omega} \left( e^{-\mathrm{i} b\Omega} e^{\mathrm{i} \omega T}\psi\right)(t)= e^{\mathrm{i} \frac{b\omega}{2}}\left(e^{\mathrm{i} ( \omega T -b \Omega )}\psi\right)(t)
 \, .
\end{equation}
In the above appears  the Weyl 
or displacement operator up to a phase factor 
\begin{equation}
\label{JPGGabWH3}
e^{\mathrm{i} ( \omega T -b \Omega)}= \mathcal{D}^{G}(b,\omega)\, , \quad \psi_{b,\omega}(t) = e^{\mathrm{i} \frac{b\omega}{2}}  \left(\mathcal{D}^{G}(b,\omega)\psi\right)(t)
 \, .
\end{equation}
The appearance of this phase factor like that one appearing in the composition formula \eqref{JPGWeylrel} indicates that 
the map $(b,\omega) \mapsto \mathcal{D}^{G}(b,\omega)$ is a \textit{projective}
representation of the time-frequency abelian plane. Dealing with a true representation necessitates the introduction of a third degree of freedom to account for this phase factor.  Hence we are led to work with the Weyl-Heisenberg group $\mathrm{G}_{\rm WH}$.   
\begin{equation}
\label{JPGWHgroupGab}
 G_{\mathrm{WH}}= \{ g = (\varsigma , b,\omega)\, , \, \varsigma \in {\mathbb R}\,,\,  (b,\omega)  \in  {\mathbb R}^{2}\} \, ,
\end{equation} 
with neutral element: $(0,0,0)$, and
\begin{equation}
\label{JPGWHGablaw}
g_{1}g_{2} = \left(\varsigma_{1} + \varsigma_{2} + \frac 1{2} (\omega_{1}b_{2} - \omega_2b_{1})\, ,   
        b_1 + b_2   ,\; \omega_{1}+\omega_{2}\right) \, , \quad g^{-1}= (-\varsigma,-b, -\omega)\,. 
\end{equation}
 The Weyl-Heisenberg group symmetry underlying the Gabor transform 
 is understood through its unitary irreducible representation (UIR). As a result of the von-Neumann uniqueness theorem, any infinite-dimensional UIR, $U$, of $\mathrm{G}_{\rm WH}$ is characterized by a real number $\lambda \neq 0$
(there is also the degenerate, one-dimensional, UIR corresponding to $\lambda = 0$).
 If the  Hilbert space carrying the UIR is the space of finite-energy signals $\mathcal{H} = L^{2}({\mathbb R},\mathrm{d}t)$, the representation operators  are defined by the action similar to \eqref{JPGGabWH1} (with the choice $\lambda = 1$) and completed with a phase factor:
\begin{equation}
\label{WHUIREP1}
U(\varsigma , b, \omega)= e^{\mathrm{i} \varsigma}
    e^{-\mathrm{i} \omega b/2} \, e^{\mathrm{i} \omega T}\, e^{-\mathrm{i} b \Omega} = e^{\mathrm{i} \varsigma}\, \mathcal{D}^G(b,\omega)\,. 
\end{equation}
With this material, it is easy to prove the Weyl-Heisenberg covariance of the Gabor transform. 
 \begin{align}
 \label{JPGWHGabcov}
\nonumber S[U(0 , b_0, \omega_0)s](b,\omega ) &= \langle \psi_{b,\omega} | U(0 , b_0, \omega_0)s \rangle
=  \langle U(0 , -b_0, -\omega_0)\,U\left(  \frac{\omega\,b}{2}, b, \omega\right)\psi | s \rangle\\
\nonumber &= \left\langle U\left(\frac{\omega\,b}{2} -\frac 1{2} (\omega_{0}b - \omega b_0),b -b_0, \omega -\omega_0\right)\psi \right |\left. s \right\rangle \\ &= e^{\mathrm{i} (\omega-\omega_{0}/2)b_0 }\, S[s](b -b_0, \omega -\omega_0)\,.
\end{align}   
 We now pick a bounded traceclass operator  $\mathfrak{Q}_0$ on $\mathcal{H}$. Its  unitary Weyl-Heisenberg transport yields the continuous family of 
bounded traceclass operators
\begin{equation}
\label{ }
\mathfrak{Q}(b,\omega)= U(\varsigma,b,\omega)\mathfrak{Q}_0 U(\varsigma,b,\omega)^{\dag}=  \mathcal{D}^G(b,\omega)\mathfrak{Q}_0  \mathcal{D}^G(b,\omega)^{\dag}\,.
\end{equation}
Applying the  Schur  Lemma  to the irreducible projective unitary representation $(b,\omega) \mapsto \mathcal{D}^G(b,\omega)$ allows to prove 
 the resolution of the identity obeyed by the operator-valued function  $\mathfrak{Q}(b,\omega)$ on the time-frequency plane 
\begin{equation}
\label{ }
\int_{\mathbb{R}^2}  \mathfrak{Q}(b,\omega)\, \frac{\mathrm{d} b \,\mathrm{d} \omega}{2\pi} = \mathbbm{1}\,. 
\end{equation}
It ensues the Weyl-Heisenberg covariant integral quantization in its most general formulation:
\begin{equation}  
\label{JPGWHCIQ1}
f(b,\omega) \mapsto A_f= \int_{\mathbb{R}^2} f(b,\omega)\, \mathfrak{Q}(b,\omega)\, \frac{\mathrm{d} b\, \mathrm{d}\omega}{2\pi}\,.
\end{equation}
The Gabor quantization corresponds to the choice $\mathfrak{Q}_0= |\psi\rangle\langle \psi|$. 
 There exists an  equivalent form of  the quantization \eqref{JPGWHCIQ1} which is expressed in terms of $\mathcal{D}^G$, 
 the Weyl transform of the operator $\mathfrak{Q}_0$ defined as the ``apodization'' function on the time-frequency plane
 \begin{equation}
\label{JPGWTQ}
\Pi(b,\omega) := \mathrm{Tr}\left(U(-b,-\omega)\mathfrak{Q}_0 \right)\,,
\end{equation}
and the symplectic Fourier transform of $f(b,\omega)$,
\begin{equation}
\label{JPGSFT}
\mathfrak{F_s}[f](b,\omega):= \int_{\mathbb{R}^2}e^{-\mathrm{i} (b\omega^{\prime}-b^{\prime}\omega)}\, f(b^{\prime},\omega^{\prime})\,\frac{\mathrm{d} b^{\prime}\,\mathrm{b} \omega^{\prime} }{2\pi}\,. 
\end{equation}
One obtains
 \begin{equation*} 
A_f= \int_{\mathbb{R}^2}  U(b,\omega)\,  \overline{\mathfrak{F_s}}[f](b,\omega)\, \Pi(b,\omega) \,\frac{\mathrm{d} b\,\mathrm{d}\omega }{2\pi } \,, \quad \Pi(b,\omega) = \mathrm{Tr}\left(U(-b,-\omega)\mathfrak{Q}_0 \right)\,.
\end{equation*}
The semi-classical portrait of $A_f$ reads:
\begin{equation*}  \begin{split}
\check f(b,\omega)  &= \int_{\mathbb{R}^2}  \left[\Pi\,\widetilde\Pi\right](b^{\prime}-b, \omega^{\prime}-\omega)\, f(b^{\prime},\omega^{\prime}) \,\frac{\mathrm{d} b^{\prime}\,\mathrm{d} \omega^{\prime}}{2\pi}  \\
&=\int_{\mathbb{R}^2}  \mathfrak{F_s}\left[\Pi\right]\ast\mathfrak{F_s}\left[\widetilde\Pi\right](b^{\prime}-b, \omega^{\prime}-\omega)\, f(b^{\prime},\omega^{\prime}) \,\frac{\mathrm{d} t^{\prime}\,\mathrm{d} \omega^{\prime}}{4\pi^2} \,,
\end{split}
\end{equation*}
where $\widetilde\Pi(b,\omega)= \Pi(-b,-\omega)$.
With a true probabilistic content,  the meaning of the convolution
\begin{equation*}
\label{JPGtruedist}
 \mathfrak{F_s}\left[\Pi\right]\ast\mathfrak{F_s}\left[\widetilde\Pi\right]
\end{equation*}
is clear: it is the probability distribution  for the difference of two vectors in the time-frequency plane, viewed as independent random variables,  and thus is adapted to the abelian and homogeneous structure of the latter (choice of origin  is arbitrary!). In a certain sense the   function $\Pi$ corresponds to the Cohen ``$f$'' function \cite{JPGcohen66} (for more details see \cite{JPGcohenbook12} and references therein) or to Agarwal-Wolf filter functions \cite{JPGagawo70}.  The simplest choice is   $\Pi(b,\omega) = 1$, of course.  Then $\mathfrak{Q}_0 = 2 \mathrm{P}$, where $\mathrm{P}$ is the parity operator. This no filtering choice yields the popular Weyl-Wigner integral quantization equivalent to the standard ($\sim$ canonical) quantization. Another example is the Born-Jordan weight,  $\Pi(b,\omega) = \dfrac{\sin b\omega}{b\omega}$, which presents appealing aspects \cite{JPGdegosson16}.

\section{Affine  quantization}
\label{JPGaffiq}
Like the Weyl-Heisenberg symmetry underlies the properties of the Gabor transform and quantization,  the affine symmetry of the open upper half plane $\Pi_{+}:=\{(b,a)\,|\, p\in\mathbb{R}\,,\, a>0\}$ explains the properties of the  continuous wavelet transform \eqref{JPGCWT2} and the resulting quantization. Together with 
\begin{enumerate}
  \item[(i)] the multiplication law
\begin{equation*}
\label{multaff}
(b_1,a_1)(b_{2},a_{2})=\left(b_1+\frac{b_2}{a_1}, a_1\,a_2 \right)\,  , 
\end{equation*}
  \item[(ii)] the unity $(0,1)$
  \item[(iii)] and the inverse
\begin{equation*}
\label{invaff}
(b,a)^{-1}= \left( - ab, \dfrac{1}{a}\right)\, , 
\end{equation*}
\end{enumerate}
 $\Pi_{+}$ is viewed as the affine group Aff$_{+}(\mathbb{R})$ of
the real line
\begin{equation*}
\mathbb{R} \ni x \overset{(b,a)}{\mapsto} (b,a)\cdot x = b + \frac{x}{a}\,.
\end{equation*}
Note that we adopt here  a definition for the dilation  which is the inverse of the standard  one used in Subsection \ref{JPGTST}. It conveniently allows to get rid of the factor $1/a^2$ present in the Lobatchevskian measure in \eqref{JPGCWT1} and \eqref{JPGCWT3}.  With the above definitions, the  measure on Aff$_{+}(\mathbb{R})$  which  is left-invariant with respect to its internal law reads   $\mathrm{d}a\,\mathrm{d}b$, i.e. is canonical. 

The affine group Aff$_{+}(\mathbb{R})$ has two non-equivalent UIR $U_{\pm}$. Both are square integrable and  this is the rationale behind continuous
wavelet analysis resulting from a resolution of the identity \eqref{JPGCWT3}. $U_{\pm}$ and the UR $ U= U_{+} \oplus U_{-}$
are realized in the Hilbert space $\mathcal{H}=L^{2}(\mathbb{R},\mathrm{d}t)= \mathrm{H}_+ \oplus  \mathrm{H}_-$, where $\mathrm{H}_\pm$ are the  (Hardy) subspaces of finite energy signals with positive and negative frequencies respectively. Here we restrict our choice to $U_{+}$, which is more conveniently realized in the present context  through the Fourier transform of signals with positive frequencies $\omega \equiv x >0$.  Its action is defined as
\begin{equation*}
L^{2}(\mathbb{R}_{+}^{\ast},\mathrm{d}x)\ni \phi(x) \mapsto U_+(b,a)\phi(x)=\frac{e^{\mathsf{i} b x}}{\sqrt{a}}\,\phi\left(\frac{x}{a}\right)\,.
\end{equation*}
Following the approach developed in Ref. \cite{JPGmur16}, we consider the operator
\begin{equation}
\label{JPGMom}
\mathsf{M}^{\varpi}:=\int_{\mathbb{R} \times \mathbb{R}_{+}^{\ast}}\mathsf{C}_{\mathrm{DM}}^{-1} U_+(b,a)\mathsf{C}_{\mathrm{DM}}^{-1} \,\varpi(b,a)\,\mathrm{d}b\,\mathrm{d}a\,, \quad\mathsf{C}_{\mathrm{DM}}\phi(x):= \sqrt{\frac{2\pi}{x}}\phi(x)\,,
\end{equation}
 with the  set of assumptions:
\begin{enumerate}
  \item[(i)] The weight function $\varpi(b,a)$ is $C^{\infty}$ on $\Pi_+$.
  \item[(ii)] It defines a tempered distribution with respect to the variable $b$ for all $a>0$. 
 \item[(iii)]  The operator $\mathsf{M}^{\varpi}$ is  self-adjoint bounded on $L^{2}(\mathbb{R}_{+}^{\ast},\mathrm{d}x)$.
\end{enumerate}
With these assumptions,  the action of $\mathsf{M}^{\varpi}$ on $\phi$ in  $L^{2}(\mathbb{R}_{+}^{\ast},\mathrm{d}x)$   is given in the form of the linear integral  operator
\begin{equation}
\label{JPGacMom}
(\mathsf{M}^{\varpi} \phi)(x) = \int_{0}^{\infty}\mathcal{M}^{\varpi}(x,x^{\prime})\,\phi(x^{\prime})\,\mathrm{d} x^{\prime}\, . 
\end{equation}
Its kernel $\mathcal{M}^{\varpi}$ is given by
\begin{equation}
\label{kerMom}
\mathcal{M}^{\varpi}(x,x^{\prime}) = \frac{1}{\sqrt{2\pi}}\,\frac{x}{x^{\prime}}\,\widehat{\varpi}_p\left( -x,\frac{x}{x^{\prime}}\right)\, . 
\end{equation}
In the above $\widehat{\varpi}_p$ is the partial Fourier transform of $\varpi$ with respect to the variable $b$: 
\begin{equation}
\label{parcoure} \widehat{\varpi}_p(y,a)=
\frac{1}{\sqrt{2\pi}}\int_{-\infty}^{+\infty} e^{-\mathrm{i} by} \varpi(b,a)\,\mathrm{d}b\, .
\end{equation}
From Schur's Lemma, one easily proves that the  affine transport of $\mathsf{M}^{\varpi}$ resolves the identity:
\begin{equation}
\label{JPGresunt}
\mathbbm{1}= \int_{\mathbb{R} \times \mathbb{R}_{+}^{\ast}}\frac{\mathrm{d} b\,\mathrm{d} a}{c_{\mathsf{M}^{\varpi}}} \,  \mathsf{M}^{\varpi}(b,a)\, ,   \quad \mathsf{M}^{\varpi}(b,a):= U_+(b,a) \mathsf{M}^{\varpi}U^{\dag}_+(b,a)\, ,
\end{equation}
provided that $0<c_{\mathsf{M}^{\varpi}} :=  \sqrt{2\pi} \int_0^{+\infty}\frac{\mathrm{d} a}{a}\,\widehat\varpi_b\left(1,-a\right)< \infty$.
The particular case \eqref{JPGCWT3} holding for the continuous wavelet transform with probe $\psi$ corresponds to  the  weight function whose partial Fourier transform is
\begin{equation}
\label{JPGacsvap}
\widehat{\varpi}_p(y,a)= \sqrt{2\pi}\, \frac{1}{a}\, \psi(-y)\,\overline{\psi\left(-\frac{y}{a}\right)}\, , \quad a>0\, , y<0\,. 
\end{equation}

It results from \eqref{JPGresunt} the affine covariant  integral  quantization of a function or distribution on the half-plane:
\begin{equation*}
f(b,a) \mapsto A^{\varpi}_f= \int_{\mathbb{R} \times \mathbb{R}_{+}^{\ast}}\frac{\mathrm{d} b\,\mathrm{d} a}{c_{\mathsf{M}^{\varpi}}} \, f(b,a)\, \mathsf{M}^{\varpi}(b,a)\, .
\end{equation*}
This map  is covariant with respect to the unitary
affine action $U_+$:
\begin{equation}
\label{JPGcovaff} U_+(b_0,a_0) A^\varpi_f U_+^{\dag}(b_0,a_0) =
A^\varpi_{\mathfrak{U}(b_0,a_0)f}\, ,
\end{equation}
with
\begin{equation}
\label{JPGcovaff2}
 \left(\mathfrak{U}(b_0,a_0)f\right)(b,a)=
f\left((b_0,a_0)^{-1}(b,a)\right)= f\left(a_0(b
-b_0),\frac{a}{a_0}, \right)\, ,
\end{equation}
$\mathfrak{U}$ being the left regular representation of the affine
group when $f\in L^2(\Pi_+, \mathrm{d} b\, \mathrm{d}a)$.

The action of $A^{\varpi}$ on $\phi$ in $C_0^{\infty}(\mathbb{R}_+^{\ast})$ is given in the form of the linear integral  operator
\begin{equation}
\label{JPGacMom}
(A^{\varpi}_f\phi)(x) = \int_{0}^{+\infty}\mathcal{A}^{\varpi}_f(x,x^{\prime})\,\phi(x^{\prime})\,\mathrm{d} x^{\prime}\, , 
\end{equation}
with kernel 
\begin{equation}
\label{JPGkerMomA}
\mathcal{A}^{\varpi}_f(x,x^{\prime}) = \frac{1}{c_{\mathsf{M}^{\varpi}}}\, \frac{x}{x^{\prime}}\int_0^{+\infty}\frac{\mathrm{d} q}{q}\,\widehat\varpi_p\left(-q,\frac{x}{x^{\prime}}\right)\, \hat{f}_p\left(x^{\prime}-x, \frac{x}{q}\right)\, .
\end{equation}
For the quantisation of the variables $b$ and $a$ (almost) nothing basic  is lost: $(A^{\varpi}_b\phi)(x) = -\mathsf{i} \partial_{x}\phi(x) + \mathrm{Cst}_3$, and  $(A^{\varpi}_a\phi)(x)=\mathrm{Cst}_4 \,x \,\phi(x)$. With an appropriate choice of the weight function $\varpi$ one gets the values  $\mathrm{Cst}_3=0$ and $\mathrm{Cst}_4=1$, and so $[A^{\varpi}_a,A^{\varpi}_b] = \mathrm{i}\mathbbm{1}$. Since $A^{\varpi}_a$ is bounded below,  $A^{\varpi}_b$, although  symmetric, is not self-adjoint and has no self-adjoint extension.

\section{Examples of operatorial signal analysis through Gabor quantization }
\label{JPGopsa}
We illustrate the content of this contribution with examples showing the link between standard operatorial tools of Signal analysis, like filtering, multiplication, convolution, etc,  and Gabor quantization 
\eqref{JPGgabq1}.  For a given choice of the probe $\psi$ the latter maps a function (or tempered distribution) $f(b,\omega)$ on the time-frequency plane to the integral operator $A_f$ in the Hilbert space of finite-energy signals defined by
\begin{equation}
\label{ }
(A_f s)(t) = \langle \delta_t  | A_f | s \rangle = \int_{-\infty}^{+\infty}\mathrm{d} t^{\prime}\, \mathcal{A}_f(t,t^{\prime})\, s(t^{\prime})\, , 
\end{equation}
with integral kernel given by
\begin{equation}
\label{ }
 \mathcal{A}_f(t,t^{\prime})= \frac{1}{\sqrt{2\pi}}\int_{-\infty}^{+\infty}\mathrm{d} b\, \widehat{f}_{\omega}(b,t^{\prime}-t)\, \psi(t-b)\,\overline{\psi(t^{\prime}-b)}
\end{equation}
Here $\widehat{f}_{\omega}(b,y)$ is the partial Fourier transform with respect to the variable $\omega$:
\begin{equation}
\label{ }
\widehat{f}_{\omega}(b,y)= \frac{1}{\sqrt{2\pi}}\int_{-\infty}^{+\infty}\mathrm{d} \omega\, f(b,\omega) \, e^{-\mathsf{i} \omega y}\,.
\end{equation}
 Let us go through some specific situations. The Gabor quantization of separable functions $f(b,\omega)= u(b)v(\omega)$ yields the integral kernel
\begin{equation}
 \mathcal{A}_{uv}(t,t^{\prime})= \frac{\hat v(t^{\prime}-t)}{\sqrt{2\pi}}\int_{-\infty}^{+\infty}\mathrm{d} b\,u(b)\, \psi(t-b)\,\overline{\psi(t^{\prime}-b)}\,,
\end{equation}
and so the action on a signal $s(t)$ reads as the combination of convolution and multiplication
\begin{equation}
(A_{uv} s)(t) =\frac{1}{\sqrt{2\pi}} \int_{-\infty}^{+\infty}\mathrm{d} b \, \psi(b) \,u(t-b) \, \left( \overline{\tilde\psi}_b \hat{\tilde v}\ast s\right)(t)\, , \quad  \psi_b(t):=\psi(t-b)\,.
\end{equation}
Therefore, in the monovariable case $f(b,\omega)= u(b)$ one gets the multiplication operator
\begin{align}
 \mathcal{A}_{u}(t,t^{\prime})&= \delta(t^{\prime}-t)\int_{-\infty}^{+\infty}\mathrm{d} b\,u(b)\, \vert\psi(t-b)\vert^2= \delta(t^{\prime}-t)\, \left(\vert \psi\vert^2\ast u\right)(t)\\
 (A_{u(b)} s)(t)&= \left(\vert \psi\vert^2\ast u\right)(t)\,s(t)\,.
\end{align}
For  the other monovariable case $f(b,\omega)= v(\omega)$ one gets  the integral kernel 
\begin{equation}
\label{JPGgabqv}
 \mathcal{A}_{v(\omega)}(t,t^{\prime})= \frac{\hat v(t^{\prime}-t)}{\sqrt{2\pi}}\int_{-\infty}^{+\infty}\mathrm{d} b\, \psi(b)\,\overline{\tilde\psi}(t-t^{\prime}-b)=  \frac{\hat v(t^{\prime}-t)}{\sqrt{2\pi}}\, R_{\psi\psi}(t-t^{\prime})
 \end{equation}
 where 
 \begin{equation}
\label{ }
R_{\psi\psi}(t):= \int_{-\infty}^{+\infty}\mathrm{d} t^{\prime}\,\psi(t^{\prime})\,\overline{\psi(t^{\prime}-t)}=\left(\psi \ast \overline{\tilde{\psi}}\right)(t)=\overline{\widetilde{R}}_{\psi\psi}(t)
\end{equation}
  is the autocorrelation of the probe,  i.e. the  correlation of the probe with a delayed copy of itself as a function of delay.
   Note that 
\begin{equation}
\frac{1}{\sqrt{2\pi}}R_{\psi\psi}(t)= \mathcal{F}^{-1}\left[\vert\hat \psi\vert^2\right](t)\,.
\end{equation}
We eventually get the  convolution operator  on the signal:
 \begin{equation}
\label{ }
 (A_{v(\omega)} s)(t)= \frac{1}{\sqrt{2\pi}} \left[\left(R_{\psi\psi}\hat{\tilde v}\right)\ast s\right](t)
\end{equation}
It is amusing to explore the above quantization formulae when the signal itself or its Fourier transform are quantized. 
Precisely,  given a normalised probe $\psi$ and a signal $s(t)$, what is the operator $A_s$, i.e., the Gabor quantisation of the signal itself?
It is given by :
\begin{equation}
\label{}
\left(A_{s(b)} s\right)(t) = \left(\vert \psi\vert^2\ast s\right)(t)\,s(t)\,.
\end{equation}
This means a kind a self-control of the original signal with its regularised version yielded by the convolution with the probability distribution $\vert \widetilde{\psi}\vert^2$.
In turn,   its Fourier transform $\hat s(\omega)$, what is the operator $A_{\hat s}$?
Applying \eqref{JPGgabqv} the Gabor quantization of the  Fourier transform $\hat s(\omega)$ of the signal  yields the convolution operator:
\begin{equation}
\left(A_{\hat s} s\right)(t) = \frac{1}{\sqrt{2\pi}} \left[\left(R_{\psi\psi}\,s\right)\ast s\right](t)\,.
\end{equation}
It is a kind of an autocorrelation  of the original signal with one of its regularised version yielded by a superposition of multiplication operators. 
 It is actually the Gabor quantization of $\overline{\widetilde{\hat s}}(\omega)$ which yields the autocorrelation of the signal weighted by  the autocorrelation of the probe 
\begin{equation}
\label{}
\left(A_{\overline{\widetilde{\hat s}}} \,s\right)(t) = \frac{1}{\sqrt{2\pi}} \int_{-\infty}^{+\infty}\mathrm{d} t^{\prime}\,R_{\psi\psi}(t^{\prime})\,s(t^{\prime})\,\overline{s(t^{\prime}-t)}\,.
\end{equation}

It is equally inspiring to Gabor quantize the Gabor transform of the signal $s$
\begin{equation}
\label{}
S(b,\omega ) =  
\int_{-\infty}^{+ \infty}
e^{-\mathsf{i}\omega t}\,\overline{\psi (t -b)} \, s(t) \, \mathrm{d} t\,.
\end{equation}
On obtains  the (involved) convolution:
\begin{equation}
\label{}
\left(A_S s\right)(t) = \int_{-\infty}^{+ \infty}\mathrm{d}b\, \psi(t-b) \,\left[\left(\overline{\psi_{b}}\,s\right)\ast\left(\overline{\psi_{-b}}\,s\right)\right](t)\,, \quad  \psi_b(t)=\psi(t-b)\,.
\end{equation}
It is worthy to examine all these formulae with the most immediate probe choice, namely the normalised centred  Gaussian with width $\sigma$ 
\begin{equation}
\label{ }
\psi(t) \equiv G_{\sigma}(t)=\frac{1}{\pi^{1/4}\sqrt{\sigma}}e^{-\dfrac{t^2}{2\sigma^2}}\,.
\end{equation} 
Its autocorrelation is also a (not normalised) Gaussian 
\begin{equation}
\label{JPGautocorG}
R_{G_{\sigma}G_{\sigma}}(t)= e^{-\frac{t^2}{4\sigma^2}}\,.
\end{equation}
The integral kernel of the quantization of $f(b,\omega)$ reads
\begin{align}
\label{ }
 \mathcal{A}_f(t,t^{\prime})&= \frac{1}{\sqrt{2}\pi\sigma}e^{-\dfrac{(t-t^{\prime})^2}{4\sigma^2}}\int_{-\infty}^{+\infty}\mathrm{d} b\,  \widehat{f}_{\omega}(b,t^{\prime}-t)\, 
 e^{-\dfrac{(b-(t+t^{\prime})/2)^2}{\sigma^2}}\\
 &= \frac{1}{\sqrt{2}\pi\sigma}e^{-\dfrac{(t-t^{\prime})^2}{4\sigma^2}} \left(\widehat{f}_{\omega}(\cdot,t^{\prime}-t)\ast e^{-\dfrac{(\cdot)^2}{\sigma^2}}\right)\left( \frac{t+t^{\prime}}{2}\right)\,.
\end{align}
Finally, note that the resulting semi-classical portrait of the operator $A_f$ is  the double Gaussian convolution: 
\begin{equation}
\label{ }
\check f (b,\omega)= \int_{\mathbb{R}^2}\frac{\mathrm{d} b^{\prime}\,\mathrm{d} \omega^{\prime}}{2\pi}\, f(b^{\prime},\omega^{\prime})\, e^{-\dfrac{(b-b^{\prime})^2}{2\sigma^2}}\,e^{-\dfrac{\sigma^2(\omega-\omega^{\prime})^2}{2}}\,.
\end{equation}
As a consequence we observe that  no classical limit holds at $\sigma \to 0$ or $\sigma \to \infty$. This is just an illustration of the time-frequency uncertainty  principle.

\section{Discussion}
\label{JPGdiscussion}

As was illustrated in the above section, it can be profitable to  view any linear operator used in signal processing, e.g., convolution, ``quantization'', compression, etc, as the quantum version of some classical $f(b,\omega)$ or  $f(b,a)$. A tentatively  complete conversion table is being established (doctoral program of C. Habonimana). New tools of signal analysis can be established in this way.

 Now,  in the quantum framework derived from Hilbertian signal analysis, measured (set of data!)  finite energy signal $s(t)$
 (resp. $\hat s(\omega)$) become ``quantum states" or ``wave functions'' and can be given a probabilistic  interpretation of some significance, \textit{e.g.}, localisation measurement in time (resp. frequency or scale).

  Hence, for a given signal $s$, in some experiment or trial, for some $f(t,\omega)$ or  $f(t,a)$, how to  interpret the ``expected values'' $\langle s|A_f|s\rangle$, e.g. $\langle s|T|s\rangle$ (resp. $\langle s|\Omega|s\rangle$)? 

 This mean time (resp. mean frequency) represents a kind of  characteristic   date (resp. frequency) for a phenomenon (e.g. an earthquake, a sound, ...) encoded into $s$. 

 It is tempting to consider the class of (deterministic or not) signals as eigenstates of some operator $A_f$ resulting from the integral quantization of a  function or distribution $f(t,\omega)$, like $\Omega e^{\mathsf{i} \omega t} = \omega e^{\mathsf{i} \omega t} $. ...

The last but not the least point, one is not familiar with considering classical electromagnetic fields or waves  as quantum states,  on which act operators which are built from  functions or distributions $f(x,k)$ on the $8$-dimensional phase space \textit{time - space $\times$ frequency - wave vector}  along the quantization procedure exposed here. Many features of quantum formalism, like entanglement or quantum measurement, remain to be explored within  this original framework, where the absence of the Planck constant  offers the opportunity to work with no limitation of scale.

\subsection*{Acknowledgement}
J. P. G. is indebted to the \textit{Ecole Doctorale de l'Universit\'e du Burundi (UB)} and his Director, Prof. Juma Shabani,  for hospitality and financial support. He is also indebted to the \textit{Centre International de Math\'ematiques Pures et Appliqu\'ees} (CIMPA) for financial support.


\begin{thebibliography}{99.}%





\bibitem{timeQM} J.G. Muga, R. Sala Mayato,  I.L. Egusquiza, (Eds.), Time in Quantum Mechanics,
Lecture Notes in Physics Monographs, Springer-Verlag Berlin Heidelberg (2008).

\bibitem{pauli58} J. W. Pauli, in Encylopaedia of physics, edited by S. Flugge (Springer, Berlin, 1958), Vol. 5, p. 60.


\bibitem{JPGber14} H. Bergeron and J.-P. Gazeau, \textit{Integral quantizations with two basic examples}, Ann. Phys.\ \textbf{344}, 43 (2014).

\bibitem{JPGbecuro17} H. Bergeron, E.M.F. Curado, J.-P. Gazeau, and Ligia M.C.S. Rodrigues,  \textit{Weyl-Heisenberg integral  quantization(s): a compendium}, arXiv:1703.08443, new version in progress

\bibitem{JPGmur16} J.-P. Gazeau, R. Murenzi \textit{Covariant affine integral quantization(s)}, J. Math. Phys. \textbf{57}, 052102 (2016). arXiv:1512.08274

\bibitem{JPGgaz18} J.-P.  Gazeau,  \textit{From classical to  quantum models: the regularising  r\^ole of integrals, symmetry and probabilities}, Found. Phys.  \textbf{48} 1648-1667  (2018);  arXiv:1801.02604.

\bibitem{JPGbergaz18} H. Bergeron and J.-P. Gazeau \textit{Variations \`a la  Fourier-Weyl-Wigner  on quantizations of the plane and the half-plane}, 
Entropy \textbf{20} 787-1-16  (2018).

\bibitem{JPGbeczuma18} H. Bergeron, E. Czuchry, and J.-P. Gazeau, and P. Ma\l kiewicz \textit{Integrable Toda system as a novel approximation to the anisotropy of Mixmaster},
Phys. Rev. D \textbf{ 98} 083512 (2018).

\bibitem{JPGkono19}J.-P. Gazeau,  T. Koide,  and D. Noguera \textit{Quantum Smooth Boundary Forces from Constrained Geometries},
 J. Phys. A: Math. Theor \textbf{52} 445203 (2019); arXiv:1902.07305v3 [quant-ph]

\bibitem{JPGvonneumann31} J. von Neumann, Die Eindeutigkeit der Schr\"{o}dingerschen Operatoren, \textit{Math. Ann.} \textbf{104} (1931), 570-578. 

\bibitem{JPGvonneumann32} J. von Neumann, \textit{Mathematische Grundlagen der Quantenmechanik}, Berlin, Springer,  1932.


\bibitem{JPGreedsimon75}  M. Reed and B. Simon, \textit{Methods of Modern Mathematical Physics, Vol. II: Fourier Analysis, Self-Adjointness}, Academic Press, 1975.


\bibitem {JPGcohen66} L. Cohen, Generalized phase-space distribution functions, \textit{J.
Math. Phys.} \textbf{7} (1966)  781-786 .

\bibitem {JPGcohenbook12} L. Cohen, \textit{The Weyl operator and its generalization},
Springer Science \& Business Media, 2012


\bibitem{JPGagawo70} B.S. Agarwal and E. Wolf, Calculus for Functions of Noncommuting Operators and General Phase-Space Methods in Quantum Mechanics, \textit{Phys. Rev. D} \textbf{2} (1970)
 2161; 2187; 2206. 

\bibitem {JPGdegosson16} M. de Gosson, \textit{Born-Jordan Quantization: Theory and
Applications}, Springer 2016.



\end{thebibliography}
\end{document}